\documentclass[aps,prb,twocolumn,superscriptaddress,longbibliography, amsmath,amssymb,amsfonts,citeautoscript]{revtex4-2}

\usepackage{graphicx}
\usepackage{epsf}
\usepackage{epstopdf}
\usepackage{graphicx}
\usepackage{tikz}
\usepackage{amsmath,bm,upgreek}
\usepackage{nccmath}
\usepackage[mathscr]{euscript}
\usepackage{natbib}
\usepackage{hyperref}

\graphicspath{{./figs/}}
\begin{document}
\title{Interplay of spin orbit interaction and Andreev reflection in proximized quantum dots }
\author{Bogdan R. Bu{\l}ka}
\affiliation{Institute of Molecular Physics, Polish Academy of
Sciences, ul. M. Smoluchowskiego 17, 60-179 Pozna{\'n}, Poland}
\author{Tadeusz Doma\'nski}
\affiliation{Institute of Physics, M. Curie-Sk\l{}odowska University, 20-031 Lublin, Poland}
\author{Karol I. Wysoki\'nski}
\affiliation{Institute of Physics, M. Curie-Sk\l{}odowska University, 20-031 Lublin, Poland}

\date{Received \today \hspace{5mm} }

\begin{abstract}
We investigate a hybrid device consisting of two quantum dots placed between a BCS superconductor and a semiconductor with a strong spin-orbit interaction. Assuming charge tunneling between quantum dots through spin-flip processes, we study the molecular Andreev bound states appearing in the proximized quantum dots. We show that the spin-orbit coupling splits a pair of the Andreev states into four in-gap quasiparticles. For the appropriate set of model parameters, two of these internal quasiparticle states merge, forming the zero-energy state. Under such circumstances, we obtain fully spin-polarized versions of the Majorana quasiparticles, localized on different quantum dots. This happens solely when the spin-orbit interaction is equally strong to the magnitude of crossed Andreev reflections, i.e. in the sweet spot. Otherwise, these processes are competitive, as can be observed in changes of the expectation values of the corresponding order parameters. We analyze signatures of such competition manifested under the nonequilibrium conditions for various configurations of the bias voltage. In particular, for the symmetric bias voltage between the normal electrodes and the Cooper pair splitter bias configuration, we reveal duality in the transport properties.  Charge transport through the zero-energy state at the sweet spot is contributed by perfectly entangled electrons with an (almost) ideal transmission. Transport studies would thus enable empirical detection of the molecular quasiparticle states and the efficiency of dissipation processes caused by the external normal electrodes.
\end{abstract}

\maketitle

\section{Introduction}

Recently, substantial progress has been made in developing quantum technology devices, in particular comprising superconductor-semiconductor hybrid structures that combine quantum coherence phenomena with experimentally tunable features of semiconductors. Such systems integrated into a circuit quantum electrodynamics ~\cite{Blais2021} enable efficient readout and manipulation of superconducting qubits~\cite{Aguado2020}. Unique properties of these hybrid structures stem from the proximity effect, inducing the Andreev bound states (ABS) on interfaces of superconductors~\cite{Likharev1979,Beenakker1991,Cuevas1996} and in the subgap region of the quantum dots attached to superconductors~\cite{Fazio1998,Kang1998}.
Especially the molecular realisations of such in-gap bound states have attracted considerable interests, due to their pivotal role in the pursuit of topological
superconductivity. The simplest platform for realisation of the molecular bound states is possible in two quantum dots proximized to a bulk superconductor. Signatures of such molecular in-gap states have been evidenced  in tunneling through semiconducting nanowires
\cite{Sherman.2017,Su.2017, Grove_Rasmussen.2018,Estrada_Saldana.2018,Pillet2019,Kornich2019,Estrada_Saldana.2020,Paaske-2020,Kurtossy2021,Zhang-2022,Junger2023,Matsuo2023,Kocsis2024,Driel2024}
 and in scanning tunneling spectroscopy for magnetic dimers deposited on superconducting samples \cite{Ruby.2018,Franke-2018,Choi.2018,Kezilebieke.2019,Sessi-2021,Yazdani_2021,Schmid2022,Franke-2024}. The Andreev bound states could be useful e.g.\ for obtaining superconducting qubits~\cite{Zazunov2003,Bretheau2013,Janvier2015}, for constructing superconducting diode~\cite{Pillet2023,Matsuo2023NP} and for {\em bottom-up} engineering of the topological superconductivity \cite{Scherubl2019,Liu2022}.

Cross-correlations between the ABSs are a necessary ingredient for the emergence of the molecular structure. Depending on the inter-dot interaction/coupling, particular properties of the molecular bound states can be obtained.
Under appropriate conditions, using superconductor-semiconductor hybrid systems with the strong spin-orbit interaction (SOI) there could appear the Majorana-type quasiparticles. Their prototype was proposed by Kitaev \cite{Kitaev2001} within the $p$-wave superconducting chain. Experimental efforts have focused so far on the semiconducting nanowires covered by conventional superconductors~\cite{Kouwenhoven2025}, planar Josephson junctions and self-organized magnetic chains on superconducting substrates \cite{Prada2020}.
Another promising approach relies on two spin-polarized quantum dots (2QD) mutually contacted through a short superconducting island~\cite{Dvir2023,Driel2024}. This {\em minimal Kitaev chain} scenario has been proposed by Leijnse and Flensberg~\cite{Leijnse2012} and later reconsidered in detail by Wimmer with coworkers \cite{Liu2022}. A pair of spatially separated Majorana bound states (MBS) could appear at fine-tuned {\em sweet spot} in the parameter space, however, their topological protection would be missing, hence the name {\em poor man’s Majorana} (PMM) states~\cite{Leijnse2012}.
The encouraging experimental data~\cite{Dvir2023} motivated Tsintzis et al. \cite{Tsintzis2022} to update the model, considering its spinful version with an additional central quantum dot embedded between two quantum dots. Such a setup allowed for better control of the crossed Andreev reflection (CAR) and the elastic cotunneling (ECT) processes in order to reach the sweet spot.
Next, Luethi et al.~\cite{Luethi2024}  derived analytical sweet spot conditions for various spinful double-dot models, by proper tuning of CAR, ECT, and local Andreev reflection (LAR) processes.
Despite intensive experimental efforts, robustness of the Majorana quasiparticles in these superconducting hybrid structures remains unclear \cite{Bordin2025}.

Appearance of the molecular Andreev states in hybrid structures with two quantum dots (2QD) proximized to an s-wave superconductor has been thoroughly investigated in the context of the Cooper pair splitter (CPS).
This setup was used to search for quantum entanglement of electrons and to test the Bell inequalities ~\cite{Burkard2000,Lesovik2001,Recher2001,Sauret2004,Burset2011,Chevallier2011,Trocha2015, Busz2017,Bulka2021}. Experimentally such a device can be realized in the electronic fork (Y-junction) geometry~\cite{Hofstetter2009,Herrmann2010,Hofstetter2011,Schindele2012,Das2012,Klobus2014,Tan2015,
Baba2018,Bordoloi2022}, where Cooper pairs arriving from the superconducting electrode were split into different quantum dots (by strong Coulomb repulsion) and transmitted to two normal electrodes. The quantum dots controlled the charge and spin transfer, whereas the normal electrodes probed the electron and spin current correlations. Furthermore, it has been proposed to integrate CPS into a waveguide circuit QED to analyze the charge susceptibility spectrum~\cite{Bulka2022}. Such integrated CPS-cQED device was recently fabricated and used to entangle the photon pairs \cite{Governale2025}.

Here, we propose to combine both approaches to form a sandwich-type structure with two quantum dots  (or in general a chain of $N$ quantum dots) embedded between s-wave superconductor and semiconductor with the strong SOI. The quantum dots should be cross-correlated via the coherent CAR and spin-flip hopping (SFH) processes, respectively. Our objective is (i) to demonstrate how these processes affect the formation of in-gap Andreev molecular states, and (ii) to establish the necessary conditions for emergence of the Majorana-type bound states. We also propose suitable empirical tools for detecting these bound states. For this purpose, we introduce two normal electrodes side-coupled to the quantum dots (see Fig.~\ref{schem}) to investigate the local/nonlocal transport properties and inspect efficiency of the decoherence processes.

The paper is organized as follows.
In Sec.~\ref{2QD} we introduce the microscopic model, describing the CAR and SFH transfer processes, which are essential for emergence of the zero-energy Majorana quasiparticles. Within the Nambu-Gorkov formalism, we next demonstrate that the CAR and the SFH terms interchange their roles in the model Hamiltonian. This duality is evident in all physical quantities. Sec.~\ref{Keldysh} presents the Keldysh Green function formalism, which enables exact calculations of the single-particle quantities treating the couplings to external electrodes, Andreev scattering, and electron spin-flip hopping on equal footing.  In Sec. \ref{spectral} we determine the quasiparticle spectrum, focusing on the interplay of CAR and SFH processes.
Since our spectrum is Dirac-type, we follow a general procedure presented in Ref.~\cite{Chamon2010} for systems describing
s-wave-induced superconductivity on the surface of topological insulators, and perform a transformation of the Green function from a local dot representation to the Majorana representation.
In Sec.~\ref{equilibrium} we calculate the thermally averaged quantities for arbitrary set of the model parameters, emphasizing duality between CAR and SFH processes.
In Sec.~\ref{currents} we present results obtained for the charge transport under various bias configurations. We derive exact analytical formulas for the transmission coefficients of our three-terminal system, based on particle transfer between the normal electrodes and the superconductor. In the sweet spot limit, these transmission coefficients reveal perfect entanglement of the transferred electrons.  Sec.~\ref{concl}  summarizes the main findings and concludes the paper.

\section{Model and Method}
\label{model}

\subsection{Microscopic model of proximized double quantum dots}
\label{2QD}

\begin{figure}
\includegraphics[width=0.9\linewidth,clip]{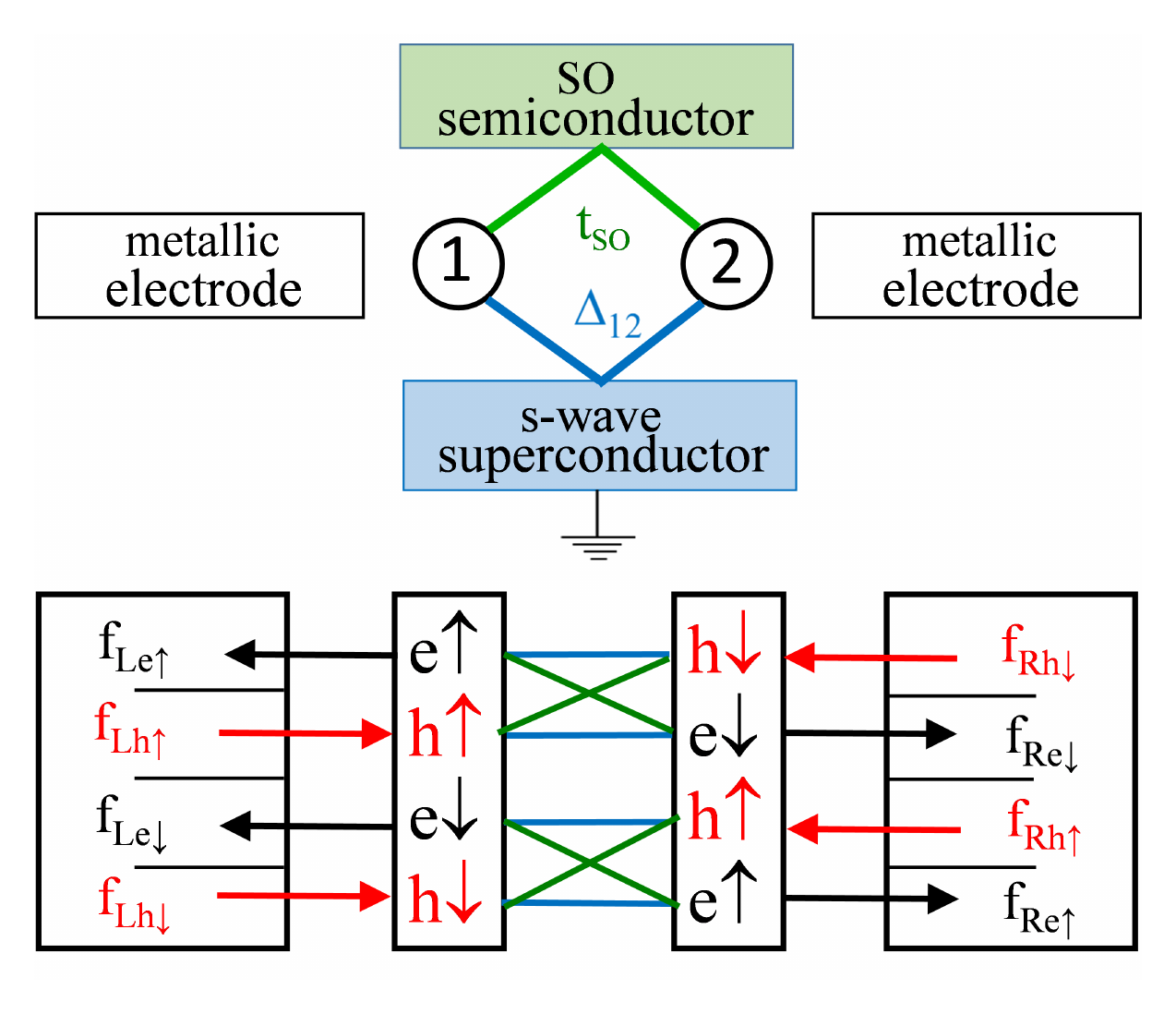}
\caption{Schematic configuration of the considered system, where two quantum dots are proximized by an s-wave superconductor and sandwiched between a semiconductor with the strong SOI.
Lower panel shows the particle and hole transfers (in black and red, respectively) between the molecule and the external metallic electrodes, which serve as the electron and hole reservoirs characterized by the Fermi distributions: $f_{Le\sigma}$, $f_{Lh\sigma}$, $f_{Re\sigma}$ and $f_{Rh\sigma}$. The inter-dot crossed Andreev reflections (CAR) are denoted in blue and the spin-flip hoppings (SFH) in green, respectively. Notice that by ignoring the local Andreev reflections and the spin-conserved hopping, we obtain two separated subspaces.
}\label{schem}
\end{figure}
Our setup consists of two quantum dots (2QDs) hybridized to an s-wave superconductor and individually coupled to the external normal electrodes, see Fig. \ref{schem}. The model Hamiltonian  can be represented by
\begin{equation}
\label{eq:ham}
H = H_{2QD}+ \sum_{\alpha=L,R,S}H_{\alpha} + H_{T} \; ,
\end{equation}
where the quantum dots are described by
\begin{align}
\label{eq:2QD}
H_{2QD} =& \sum_{i=1,2;\sigma=\uparrow,\downarrow} \epsilon_{i\sigma} c^{\dag}_{i\sigma}c_{i\sigma} + \sum_{\sigma=\uparrow,\downarrow} t (c^{\dag}_{1\sigma}c_{2\sigma}+ c^{\dag}_{2\sigma}c_{1\sigma})\nonumber\\
&+ t_{so}(c^{\dag}_{1\uparrow}c_{2\downarrow}+c^{\dag}_{2\downarrow}c_{1\uparrow}-
c^{\dag}_{1\downarrow}c_{2\uparrow}-c^{\dag}_{2\uparrow}c_{1\downarrow})\nonumber\\
&+ \sum_{i} U_{i}\; n_{i\uparrow} n_{i\downarrow}
\end{align}
The first term on r.h.s.  of Eqn.~(\ref{eq:2QD}) describes the local single energy levels $\epsilon_i$ of the i-th QD. The second term corresponds to the spin-conserved inter-dot hopping integral, $t$. In the CPS geometry, such hopping is harmful for the entanglement of the electrons, reducing the splitting efficiency. The third term describes SFH between the quantum dots, originating from the spin-orbit interaction combined with the Zeeman field~\cite{Stepanenko2012,Hussein2016,Tsintzis2022,Luethi2024,Spethmann2024}. The last term in Eqn.~(\ref{eq:2QD}) refers to the strongly repulsive on-dot Coulomb potential, $U_i$, suppressing the double occupancies of both quantum dots.

Electrons of the external $\alpha$-th electrodes are described by
\begin{align}
H_{\alpha}=\sum_{k\sigma}\varepsilon_{\alpha k} c^{\dag}_{\alpha k\sigma} c_{\alpha k\sigma}- \delta_{\alpha,S}\;\Delta\sum_{k}(c^{\dag}_{S k\uparrow}c^{\dag}_{S -k\downarrow}+\text{h.c.}) ,
\end{align}
where $\varepsilon_{\alpha k}$ is the electron energy and $\Delta$  denotes the pairing gap of superconductor. The hybridization terms
\begin{align}
H_T=\sum_{k,\sigma}(& t_{L1} c^{\dag}_{Lk\sigma}c_{1\sigma} + t_{R2} c^{\dag}_{Rk\sigma}c_{2\sigma}+ t_{S1} c^{\dag}_{Sk\sigma}c_{1\sigma}\nonumber\\
& + t_{S2} c^{\dag}_{Rk\sigma}c_{2\sigma}+\text{h.c.})
\end{align}
describe the electron hopping, $t_{\alpha i}$,  between $\alpha$-th electrode and $i$-th QD.

We focus our considerations to the subgap regime $|E| < \Delta$, thus for simplicity we impose the superconducting atomic limit $\Delta \rightarrow\infty$. In this case, the fermionic degrees of freedom of superconducting electrode can be integrated out~\cite{Rozkov2000,Meng2009}. In effect, the proximization process yields the singlet pairings~\cite{Eldridge2010,Hussein2016}    ~\cite{Eldridge2010,Hussein2016}
\begin{align}\label{eq:prox}
H_{S2QD} = & H_{2QD}-\sum_{i=1,2}\Delta_{i}(c^{\dag}_{i\uparrow}c^{\dag}_{i\downarrow}-
c_{i\uparrow}c_{i\downarrow})\nonumber\\
&-\Delta_{12}(c^{\dag}_{1\uparrow}c^{\dag}_{2\downarrow}+c^{\dag}_{2\uparrow}c^{\dag}_{1\downarrow}
-c_{1\uparrow}c_{2\downarrow}-c_{2\uparrow}c_{1\downarrow}).
\end{align}
The on-dot term describes the local Andreev reflection (LAR), where $\Delta_{i} = \pi\rho_{S}t_{Si}^2$ and $\rho_{S}$ denotes the density of states of the S-th electrode near the Fermi level in its normal state. The second (inter-dot) term corresponds to the CAR processes, where the inter-dot coupling is $\Delta_{12}= \pi\rho_{S} t_{S1} t_{S2}$.  In the CPS framework, these CAR processes are responsible for electron entanglement, whereas LAR reduce the splitter efficiency.

Let us compare the above model with other scenarios used in the literature, like~\cite{Tsintzis2022,Luethi2024}. First, we assume a sandwich-like geometry  in contrast to the quasi-one-dimensional models in previous works, where the quantum dots are contacted through superconducting island (or via the proximized central dot). We thus envisage the situation in which both quantum dots are coupled to a bulk superconductor. Furthermore (probably even more importantly), the spin-orbit interaction is assumed to operate between the quantum dots, while in Ref.~\cite{Luethi2024} it acts on interface between a superconductor (or central QD) and the neighboring dots. Source of this spin-orbit interaction is the semiconducting segment of our sandwich structure ($c.f.$ the upper panel in Fig. (\ref{schem})).

The Coulomb potential $U_i$ is usually orders of magnitude larger than the local pairing amplitude $\Delta_i$. Strong competition of the on-dot repulsion with the superconducting proximity effect makes the doubly occupied configuration of each QD hardly possible. The single occupancy is far more favorable, especially near a half-filling of the quantum dots weakly coupled to the superconducting electrode \cite{Bauer2007}. For this reason, in what follows, we discard the doubly occupied configurations so that the Coulomb potential is ineffective. Our considerations are hence relevant to the case with the singlet inter-dot pairing.

Introducing the Nambu notation $\Psi^{\dag}_{2QD}=[c^{\dag}_{1\uparrow},c_{1\uparrow},c^{\dag}_{2\downarrow},c_{2\downarrow},
c^{\dag}_{1\downarrow},c_{1\downarrow},c^{\dag}_{2\uparrow},c_{2\uparrow}]$ we can describe the proximized quantum dots by the following compact expression
\begin{align}
H_{S2QD}=\frac{1}{2}\Psi^{\dag}_{2QD}\mathcal{H}_{S2QD}\Psi_{2QD} + \text{const}.,
\end{align}
where the matrix Hamiltonian is given by
\begin{align}\label{hs2qd}
&\mathcal{H}_{S2QD}=\nonumber\\
&\left[\begin{array}{cccccccc}
 \epsilon_{1\uparrow} & 0 &t_{so}&-\Delta_{12} &  0&-\Delta_{1} & t&0 \\
0& -\epsilon_{1\uparrow}&\Delta_{12}& -t_{so}&\Delta_1 &0 &0&-t \\
t_{so}& \Delta_{12}&\epsilon_{2\downarrow}& 0& t& 0& 0& \Delta_{2}\\
-\Delta_{12}&-t_{so}&0&-\epsilon_{2\downarrow}& 0&-t&-\Delta_{2}&0\\
 0&\Delta_1 &t &0 &  \epsilon_{1\downarrow} &0&-t_{so} &\Delta_{12} \\
-\Delta_{1}&0 &0 &-t& 0&-\epsilon_{1\downarrow}& -\Delta_{12} &t_{so} \\
t&0 &0 &-\Delta_{2}&  -t_{so}& -\Delta_{12} & \epsilon_{2\uparrow}&0 \\
0 &-t &\Delta_2 &0 &  \Delta_{12}&  t_{so}& 0 & -\epsilon_{2\uparrow}
\end{array}\right].
\end{align}
Specifically, focusing on the particle-hole symmetry (PHS)  ($\epsilon_{1\sigma}=\epsilon_{2\sigma}=0$) and assuming identical on-dot pairings $\Delta_1=\Delta_2$, we obtain the eigenenergies of the matrix (\ref{hs2qd})
\begin{align}
E=\pm \sqrt{t^2+\left(\Delta_{12}\pm\sqrt{t_{so}^2+\Delta_1^2}\right)^2}.
\end{align}
We notice that the parameters $\Delta_{12}$ and $t_{so}$ play a decisive role in achieving the zero-energy state in the spectrum. This situation occurs for $t=0$, $\Delta_{12}=\pm \sqrt{t_{so}^2+\Delta_1^2}$. Upon transforming the operators of spin-$\downarrow$ electrons to the hole operators: $\{c^{\dag}_{1\downarrow}, c^{\dag}_{2\downarrow}\}\leftrightarrow \{c_{1\downarrow}, c_{2\downarrow}\} $ the CAR and SFH terms are interchanged in the Hamiltonian [see the corresponding expressions in Eqs (\ref{eq:2QD}) and (\ref{eq:prox})]. We hence claim that CAR and SFH are {\it dual} to each other.

Let us consider in more detail the special case where the hopping integral $t$ and the on-dot pairing potential $\Delta_i$ are both absent. The latter assumption can be achieved in practice by applying the spin-polarized side gates, as has been reported for this CPS geometry in Ref.~\cite{Bordoloi2022}.
Under these circumstances, the matrix Hamiltonian (\ref{hs2qd}) simplifies to a block-diagonal structure (as graphically sketched in Fig.\ref{schem})~. One block of this Hamiltonian is given by
\begin{align}\label{hsplitso}
\mathcal{H}_{\Psi}=
\left[\begin{array}{cccc}
 \epsilon_{1\uparrow} &0 & t_{so}&-\Delta_{12} \\
0& -\epsilon_{1\uparrow}&\Delta_{12}& -t_{so}\\
t_{so}& \Delta_{12}&\epsilon_{2\downarrow}& 0\\
-\Delta_{12} & -t_{so}&0& -\epsilon_{2\downarrow}
\end{array}\right]
\end{align}
in the representation $\Psi^{\dag}=[c^{\dag}_{1\uparrow},c_{1\uparrow},c^{\dag}_{2\downarrow}, c_{2\downarrow}]$. The second subspace, represented by $\Phi^{\dag}=[c^{\dag}_{1\downarrow}, c_{1\downarrow},c^{\dag}_{2\uparrow},c_{2\uparrow}]$,
is described by the following part of the matrix Hamiltonian
\begin{align}\label{hsplitsodu}
\mathcal{H}_{\Phi}=
\left[\begin{array}{cccc}
 \epsilon_{1\downarrow} &0 & -t_{so}&\Delta_{12} \\
0& -\epsilon_{1\downarrow}&-\Delta_{12}& t_{so}\\
-t_{so}& -\Delta_{12}&\epsilon_{2\uparrow}& 0\\
\Delta_{12} & t_{so}&0& -\epsilon_{2\uparrow}
\end{array}\right].
\end{align}
We further investigate only the $\mathcal{H}_{\Psi}$ subspace, because the properties of $\mathcal{H}_{\Phi}$ can be deduced by exchanging the model parameters [$\epsilon_{1\uparrow}$, $\epsilon_{2\downarrow}$, $t_{so}$, $\Delta_{12}$] $\rightarrow$ [$\epsilon_{1\downarrow}$, $\epsilon_{2\uparrow}$, $-t_{so}$, $-\Delta_{12}$].

The Hamiltonian (\ref{hsplitso}) of the $\mathcal{H}_{\Psi}$ subspace is strictly analogous to the poor man's scenario \cite{Leijnse2012}, where the authors considered the two-site Kitaev chain. In our case, however, the role of one site  is played by $\uparrow$-spin of QD$_1$ and the other site refers to $\downarrow$-spin of QD$_2$. Instead of the intersite pairing between spinless fermions, we have the inter-dot pairing $\Delta_{12}$ of opposite spin electrons. Role of the hopping integral between two sites of the Kitaev chain is played by the spin-reversal hopping, $t_{SO}$. For $\epsilon_{1\uparrow}=0=\epsilon_{2\downarrow}$ the eigenvalues of (\ref{hsplitso}) occur at $E=\pm(\Delta_{12} -t_{SO})$. It implies appearance of the zero-energy quasiparticle for the case $\Delta_{12}=t_{SO}$ exactly in the same fashion as predicted in Ref.~\cite{Leijnse2012}. In subsection \ref{spectral} we shall inspect whether this quasiparticle has the Majorana-type properties, or not.

\subsection{Keldysh Green function approach}
\label{Keldysh}

For studying the spectroscopic features of superconducting hybrid nanostructures and analyze their transport properties it is convenient to use the Green function approach \cite{Cuevas1996,Martin-Rodero2011,Sun2000}. In particular, this formalism has been applied to the proximized 2QDs coupled to the normal electrodes \cite{Chevallier2011,Dong2017,Bulka2021,Bulka2022}. Here we focus on noninteracting particles, therefore we can solve exactly the equation of motion for the nonequilibrium (Keldysh) Green functions.

This formalism applied to the $\mathcal{H}_{\Psi}$ sector, and coupled to the L and R-normal electrodes, gives the following Keldysh Green function
\begin{align}
\hat{G}_{\Psi}\equiv
\left[\begin{array}{cccc}
\hat{z}_{e1\uparrow} & 0&-\hat{t}_{so}
& \hat{\Delta}_{12} \\
0& \hat{z}_{h1\uparrow} &-\hat{\Delta}_{12} &t_{so}\\
-\hat{t}_{so} &-\hat{\Delta}_{12} &\hat{z}_{e2\downarrow}& 0\\
\hat{\Delta}_{12}&t_{so}&0&\hat{z}_{h2\downarrow}\end{array}\right]^{-1}
\label{eq:KGF}
\end{align}
expressed in the Nambu representation $\Psi^{\dag}=[c^{\dag}_{1\uparrow},c_{1\uparrow},c^{\dag}_{2\downarrow}, c_{2\downarrow}]$.
Its elements are given in the Keldysh notation
\begin{align}
&\hat{z}_{e1\uparrow} \equiv
\left[\begin{array}{cc}
z^{--}_{e1\uparrow}& z^{-+}_{e1\uparrow}\\
z^{+-}_{e1\uparrow} &z^{++}_{e1\uparrow}
\end{array}\right]=\nonumber\\
&\left[\begin{array}{cc}
\omega-\epsilon_{1\uparrow}+\imath (1 -2f_{Le})\gamma_L& 2 \imath f_{Le}\gamma_L \\
2 \imath (-1+f_{Le})\gamma_L &-\omega+\epsilon_{1\uparrow}+\imath (1 -2f_{Le})\gamma_L
\end{array}\right],\\
&\hat{z}_{h1\uparrow}=\nonumber\\
&\left[\begin{array}{cc}
\omega+\epsilon_{1\uparrow}+\imath (1 -2f_{Lh})\gamma_L& 2 \imath f_{Lh}\gamma_L \\
2 \imath (-1+f_{Lh})\gamma_L &-\omega-\epsilon_{1\uparrow}+\imath (1 -2f_{Lh})\gamma_L
\end{array}\right].
\end{align}
Similarly one can derive $\hat{z}_{e2\downarrow}$, $\hat{z}_{h2\downarrow}$ corresponding to the second quantum dot which is coupled to the R-electrode. Here, we introduced the selfenergy
\begin{eqnarray}\label{eq:normal}
\hat{\Sigma}_{Le(h)}=\imath\gamma_{L}
\left[\begin{array}{cc}
2f_{L e(h)}-1&2f_{L e(h)}\\
2(f_{L e(h)}-1)&2f_{L e(h)}-1
\end{array}\right],
\end{eqnarray}
which describes coupling of QD$_1$ to the $L$ normal electrode as a reservoir of the electrons and holes
characterized by the Fermi distribution functions $f_{L e}=\{\exp[(E-\mu_{L})/k_BT]+1\}^{-1}$ and $f_{L h}=\{\exp[(E+\mu_{L})/k_BT]+1\}^{-1}$ with an electrochemical potential $\mu_L$. The superconductor is assumed to be grounded, $\mu_S=0$. The selfenergy was derived in the wide flat-band approximation, with $\gamma_{L} = \pi\rho_{L}|t_{L}|^2$ the same for electrons and holes, where $\rho_{L}$ denotes the density of states in the $L$ electrode. Dissipation by the electron and hole reservoirs is assumed to be identical. We also used the Keldysh notation for $\hat{t}_{so}= t_{so} \hat{\tau}_z$ and $\hat{\Delta}_{12}= \Delta_{12} \hat{\tau}_z$, where $\hat{\tau}_z$ is the z-component of the Pauli matrix.
The retarded and lesser Green functions are given by $G^r_{\uparrow\downarrow}=G^{--}_{\uparrow\downarrow} -G^{-+}_{\uparrow\downarrow}$ and $G^<_{\uparrow\downarrow}=G^{-+}_{\uparrow\downarrow}$, respectively.

\section{Properties of the Andreev molecule coupled to electrodes}
\label{main}

\subsection{Quasiparticle spectrum}
\label{spectral}

Using the retarded Green function
\begin{widetext}
\begin{align}\label{retarded}
\hat{G}^r_{\Psi} (\omega)\equiv\ll \Psi|\Psi^{\dag}\gg^r_{\omega}=
\left[\begin{array}{cccc}
\ll c_{1\uparrow}|c^{\dag}_{1\uparrow}\gg^r_{\omega} &
\ll c_{1\uparrow}|c_{1\uparrow}\gg^r_{\omega}&
\ll c_{1\uparrow}|c^{\dag}_{2\downarrow}\gg^r_{\omega} &
\ll c_{1\uparrow}|c_{2\downarrow}\gg^r_{\omega}
\\
\ll c^{\dag}_{1\uparrow}|c^{\dag}_{1\uparrow}\gg^r_{\omega} &
\ll c^{\dag}_{1\uparrow}|c_{1\uparrow}\gg^r_{\omega}&
\ll c^{\dag}_{1\uparrow}|c^{\dag}_{2\downarrow}\gg^r_{\omega} &
\ll c^{\dag}_{1\uparrow}|c_{2\downarrow}\gg^r_{\omega}
\\
\ll c_{2\downarrow}|c^{\dag}_{1\uparrow}\gg^r_{\omega} &
\ll c_{2\downarrow}|c_{1\uparrow}\gg^r_{\omega}&
\ll c_{2\downarrow}|c^{\dag}_{2\downarrow}\gg^r_{\omega} &
\ll c_{2\downarrow}|c_{2\downarrow}\gg^r_{\omega}
\\
\ll c^{\dag}_{2\downarrow}|c^{\dag}_{1\uparrow}\gg^r_{\omega} &
\ll c^{\dag}_{2\downarrow}|c_{1\uparrow}\gg^r_{\omega}&
\ll c^{\dag}_{2\downarrow}|c^{\dag}_{2\downarrow}\gg^r_{\omega} &
\ll c^{\dag}_{2\downarrow}|c_{2\downarrow}\gg^r_{\omega}
\end{array}\right],
\end{align}
we can determine the spectral density, which is crucial for computing the expectation values of physical quantities in our model. We have analytically calculated all the terms of this matrix function (\ref{retarded}). In general, their form is rather complicated, therefore we will show them only for the selected special cases.
\end{widetext}

\begin{figure}
\includegraphics[width=0.9\linewidth,clip]{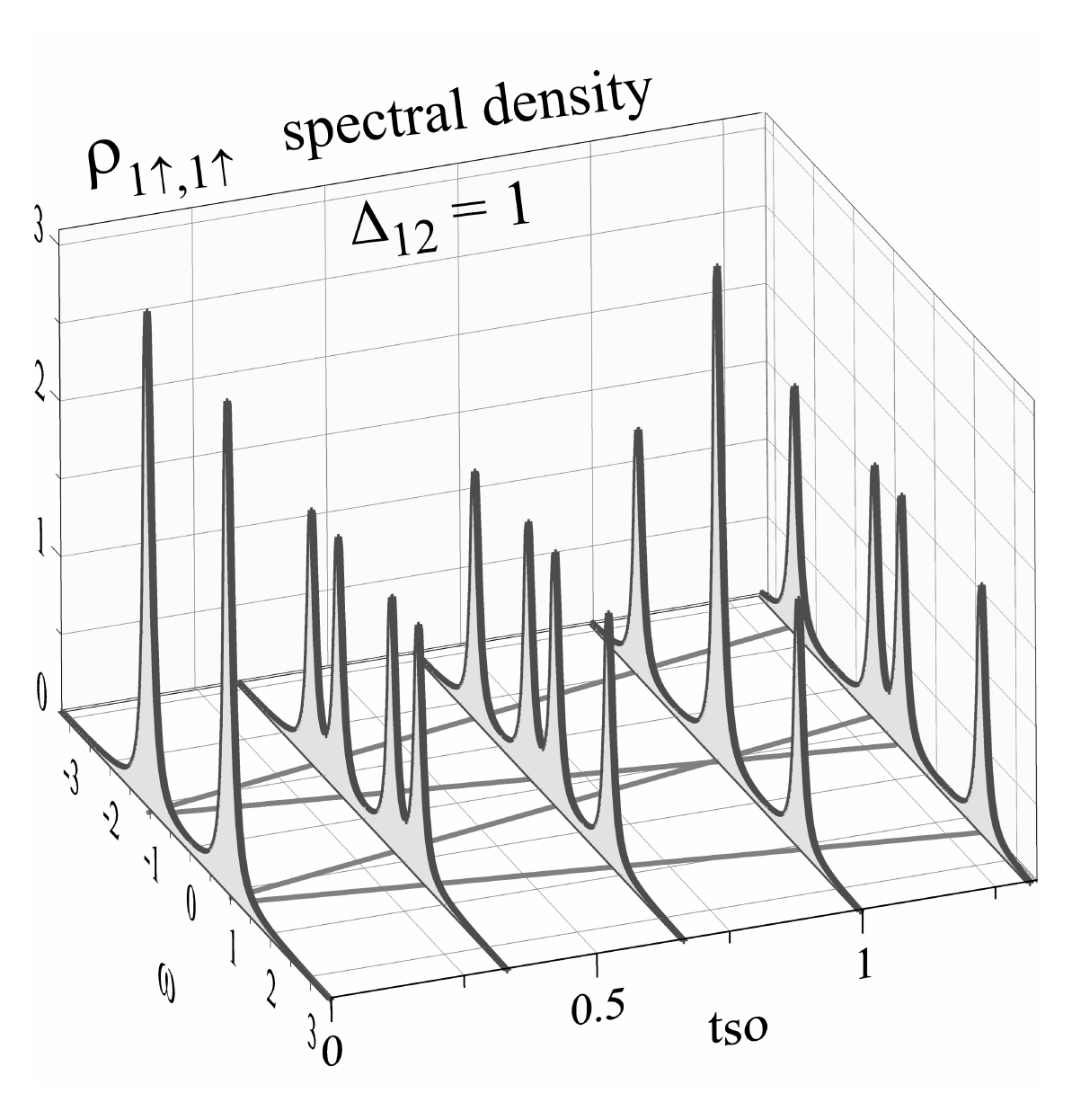}
\caption{The spectral function $\rho_{1\uparrow,1\uparrow}=(-1/\pi)\Im \ll c_{1\uparrow}|c^{\dag}_{1\uparrow}\gg^r_{\omega}$ for $\Delta_{12}=1$ and several values of the spin-orbit coupling $t_{so}=0$, $1/3$, $2/3$, $1$ and $4/3$. The horizontal lines show the positions of the poles: $\omega=\pm (\Delta_{12} \pm t_{so})$. Results are obtained for the particle-hole symmetric case $\epsilon_1=0=\epsilon_2$, assuming small symmetric couplings to the external normal electrodes $\gamma=0.1$.}\label{Fspectral}
\end{figure}

We start by studying the influence of spin-orbit coupling, $t_{SO}$, on the molecular structure of the bound states. Fig.~\ref{Fspectral} displays variation of the spectrum of the Andreev molecule with respect to the spin-orbit coupling obtained for the particle-hole symmetric case, $\epsilon_{1\uparrow}=0=\epsilon_{2\downarrow}$. In absence of the spin-orbit interaction between the dots, $t_{so}=0$, the spectrum is represented a one pair of ABS at energies $\omega=\pm E_A=\pm \Delta_{12}$. The spin-orbit interaction splits each of the Andreev bound states into  four states $\omega= \pm|\Delta_{12}\pm t_{so}|$. In particular, a pair of the internal sub-peaks merge at $\Delta_{12}=t_{so}$, forming the zero-energy quasiparticle states. The spectral weight of this central peak is doubled in comparison to the remaining states, while the total weight is conserved. For the stronger couplings $t_{so}>\Delta_{12}$ the peaks split again. The sub-peaks move in the same direction in frequency space increasing their distance.

Interestingly, exactly the same behavior is observed in evolution of the quasiparticle states of two quantum dots for the fixed spin-orbit coupling, e.g.\ $t_{so}=1$, upon increasing the pairing $\Delta_{12}$. The CAR term imposes a splitting onto the initial quasiparticles states, which is proportional to $\Delta_{12}$. The effective spectral function looks similar to the one presented in Fig.~\ref{Fspectral}. In other words, the spectral function $\rho_{1\uparrow,1\uparrow}(\omega)$ is invariant on the exchange $t_{so} \leftrightarrow \Delta_{12}$. It manifests {\em duality} between the quasiparticle states induced in the double quantum dot by the spin-orbit interaction and the molecular Andreev bound states due to the interdot pairing.

The Green $\hat{G}^r_{\Psi}$, Eq.(\ref{retarded}) expressed in the local dot representation has rather complicated structure. Therefore, we impose transformation to the Dirac representation to get better insight into physics.
At particle-hole symmetry the local Hamiltonian (\ref{hsplitso}) can be diagonalized for arbitrary $t_{so} \neq \Delta_{12}$ by the matrix
\begin{align}
S=\frac{1}{2}
\left[\begin{array}{rrrr}
-1 &1 & 1&1 \\
1& -1& 1& 1\\
1& 1 & -1& 1\\
1 &  1& 1& -1
\end{array}\right] .
\end{align}
The new eigenbasis is expressed by the Dirac fermion operator $\hat{X}^{\dag}=[a^{\dag},a,b^{\dag},b]$, where
$a^{\dag}=(-c^{\dag}_{1\uparrow}+c_{1\uparrow}+c^{\dag}_{2\downarrow}+ c_{2\downarrow})/2$ and $b^{\dag}=(c^{\dag}_{1\uparrow}+c_{1\uparrow}-c^{\dag}_{2\downarrow}+ c_{2\downarrow})/2$. Transforming the Green function (\ref{retarded}) to the Dirac basis $\hat{G}_{X}^r (\omega)= \ll \hat{X}|\hat{X}^{\dag}\gg^r_{\omega} = S^{-1} \hat{G}^r(\omega) S$ we obtain the block-diagonal structure
\begin{align}
\hat{G}_{X}^r(\omega) =
\left[
\begin{array}{cc}
\hat{G}_{A}^r(\omega)&0\\
0&\hat{G}_{B}^r(\omega)
\end{array}
\right]
\end{align}
with two separated subspaces. The first one with the Green function

\begin{widetext}
\begin{align}
\hat{G}_{A}^r(\omega) &\equiv \ll \hat{A}|\hat{A}^{\dag}\gg^r_{\omega} \nonumber\\
&=
\left[\begin{array}{cc}
\cfrac{\omega-(\Delta_{12}-t_{so})+\imath (\gamma_L+\gamma_R)/2}{(\omega +\imath \gamma_L)(\omega +\imath \gamma_R)-(\Delta_{12}- t_{so})^2}&\cfrac{\imath (\gamma_L- \gamma_R)/2}{(\omega +\imath \gamma_L)(\omega +\imath \gamma_R)-(\Delta_{12}-t_{so})^2} \\
\cfrac{\imath (\gamma_L-\gamma_R)/2}{(\omega +\imath \gamma_L)(\omega +\imath \gamma_R)-(\Delta_{12}-t_{so})^2}& \cfrac{\omega+(\Delta_{12}-t_{so})+\imath (\gamma_L+\gamma_R)/2}{(\omega +\imath \gamma_L)(\omega +\imath \gamma_R)-(\Delta_{12}-t_{so})^2}
\end{array}
\right]\label{GFA}
\end{align}
corresponds to the representation $\hat{A}^{\dag}=[a^{\dag},a]$
with a pair of the quasiparticle states at $\pm|\Delta_{12}-t_{so}|$. These states are represented by the inner peaks in the spectral function shown in Fig.~\ref{Fspectral}. For the second subspace one gets
 \begin{align}
\hat{G}_{B}^r(\omega) &\equiv \ll \hat{B}|\hat{B}^{\dag}\gg^r_{\omega} \nonumber\\
&=
\left[\begin{array}{cc}
\cfrac{\omega+(\Delta_{12}+t_{so})+\imath (\gamma_L+\gamma_R)/2}{(\omega +\imath \gamma_L)(\omega +\imath \gamma_R)-(\Delta_{12}+ t_{so})^2}&\cfrac{\imath (\gamma_R- \gamma_L)/2}{(\omega +\imath \gamma_L)(\omega +\imath \gamma_R)-(\Delta_{12}+t_{so})^2} \\
\cfrac{\imath (\gamma_R-\gamma_L)/2}{(\omega +\imath \gamma_L)(\omega +\imath \gamma_R)-(\Delta_{12}+t_{so})^2}& \cfrac{\omega-(\Delta_{12}+t_{so})+\imath (\gamma_L+\gamma_R)/2}{(\omega +\imath \gamma_L)(\omega +\imath \gamma_R)-(\Delta_{12}+t_{so})^2}
\end{array}
\right],\label{GFB}
\end{align}\end{widetext}
which refers to the representation $\hat{B}^{\dag}=[b^{\dag},b]$ with the quasiparticle states at nonzero energies $\pm|\Delta_{12}+t_{so}|$ displayed  in Fig.~\ref{Fspectral} by the outer peaks of the spectral function. For the asymmetric couplings to external electrodes, $\gamma_L\neq\gamma_R$, both Green functions (\ref{GFA})-(\ref{GFB}) have nonvanishing off-diagonal terms which are responsible for dissipation processes. This shows the importance of external reservoirs for the qualitative properties of our setup.

In particular, for $\Delta_{12}=t_{so}=g$, the inner peaks of the spectrum merge forming the zero-energy quasiparticle state. In this {\em sweet spot}, the function (\ref{GFA}) can be diagonalized by the transformation to Majorana fermions:
$\gamma_{1\uparrow}^{\dag}=\gamma_{1\uparrow}=\imath(a-a^{\dag})/\sqrt{2}=\imath(c^{\dag}_{1\uparrow} -c_{1\uparrow})/\sqrt{2}$ and $\eta^{\dag}_{2\downarrow}=\eta_{2\downarrow}=(a+a^{\dag})/\sqrt{2} =(c^{\dag}_{2\downarrow}+c_{2\downarrow})/\sqrt{2}$. In this representation, the function $\hat{G}_{A}^r$ simplifies to
\begin{widetext}
\begin{align}
\left[\begin{array}{cc}
\ll \gamma_{1\uparrow}|\gamma_{1\uparrow}^{\dag}\gg^r_{\omega} &
\ll \gamma_{1\uparrow}|\eta_{2\downarrow}^{\dag}\gg^r_{\omega} \\
\ll \eta_{2\downarrow}|\gamma_{1\uparrow}^{\dag}\gg^r_{\omega}&
\ll \eta_{2\downarrow}|\eta_{2\downarrow}^{\dag}\gg^r_{\omega}
\end{array}
\right]=\left[\begin{array}{cc}
\cfrac{1}{\omega+\imath \gamma_L}&
0\\
0&
\cfrac{1}{\omega+\imath \gamma_R}
\end{array}
\right].\label{GFAM}
\end{align}
It describes the zero-energy states existing separately on different quantum dots and in different spin sectors. Their broadening (finite lifetime) is due to the couplings to external reservoirs. We emphasize that such zero-energy quasiparticles appeared in the original {\em minimal Kitaev chain} scenario \cite{Leijnse2012}, however, here they are in the opposite spin sectors. Moreover, we obtain four Majorana modes: two of them for each spin direction. This is important generalization of the Kitaev's approach originally proposed for spinless particles.

On the other hand, the Green function of the B-sector, $\hat{G}_{B}^r$, refers to the quasiparticle states at finite energies $\omega=\pm 2 g$. Introducing the other pair of Majorana operators
$\eta_{1\uparrow}^{\dag}=\eta_{1\uparrow}=(b+b^{\dag})/\sqrt{2}=(c^{\dag}_{1\uparrow} +c_{1\uparrow})/\sqrt{2}$, and $\gamma_{2\downarrow}^{\dag}=\gamma_{2\downarrow}=\imath(b-b^{\dag})/\sqrt{2} =\imath(c^{\dag}_{2\downarrow}-c_{2\downarrow})/\sqrt{2}$ we obtain the following Green function
\begin{align}
\left[\begin{array}{cc}
\ll \eta_{1\uparrow}|\eta_{1\uparrow}^{\dag}\gg^r_{\omega}&
\ll \eta_{1\uparrow}|\gamma_{2\downarrow}^{\dag}\gg^r_{\omega} \\
\ll \gamma_{2\downarrow}|\eta_{1\uparrow}^{\dag}\gg^r_{\omega}&
\ll \gamma_{2\downarrow}|\gamma_{2\downarrow}^{\dag}\gg^r_{\omega}
\end{array}
\right]=\left[\begin{array}{cc}
\cfrac{\omega+\imath \gamma_R}{(\omega+\imath \gamma_L)(\omega+\imath \gamma_R)-4g^2}&
\cfrac{-2\imath g}{(\omega+\imath \gamma_L)(\omega+\imath \gamma_R)-4g^2} \\
\cfrac{2\imath g}{(\omega+\imath \gamma_L)(\omega+\imath \gamma_R)-4g^2}&
\cfrac{\omega+\imath \gamma_L}{(\omega+\imath \gamma_L)(\omega+\imath \gamma_R)-4g^2}
\end{array}
\right] .\label{GFBM}
\end{align}\end{widetext}
In this representation, the matrix Green function (\ref{GFBM}) has the off-diagonal terms. The corresponding finite-energy states should be interpreted as the molecular ABS, originating from the inter-dot hybridization.

Interestingly, by reversing the sign of {SFH}, $t_{so} \rightarrow -t_{so}$, the electron and hole hoppings are exchanged, and the sectors $A$ and $B$ become interchanged.
We should keep in mind that in the $\Phi$ subspace, the spins and the Majorana polarization are reversed.

In this context, one may recall the paper~\cite{Chamon2010} in which authors quantize the Majorana fermions in two-dimensional topological superconductors by noting first the similarity of equations governing excitations of proximized topological insulators to the Dirac equation. Here, we deal with zero-dimensional states, which for a particular set of parameters describe the quasi-particles, which happen to be their own anti-quasi-particles. In Sec.\ref{currents}, analyzing transmission coefficients across the structure, we will study the features of these  Majorana states detectable in charge transport.

\subsection{Thermal averages at equilibrium}\label{equilibrium}

We now study the thermal averages of various observables in our model, which can expressed by the lesser Green function
\begin{align}\label{thermal}
\left\langle \Psi^{\dag} \Psi \right\rangle=-\imath\int \frac{d\omega}{2\pi}\ll \Psi|\Psi^{\dag}\gg^<_{\omega} .
\end{align}
This relation is valid in equilibrium and non-equilibrium situations. At equilibrium, Eq.~(\ref{thermal}) simplifies,  because the lesser Green function is given by $\ll \Psi|\Psi^{\dag}\gg^<_{\omega}=\ll \Psi|\Psi^{\dag}\gg^{-+}_{\omega} = -2\imath f(\omega) \Im [ \ll \Psi| \Psi^{\dag}\gg^r_{\omega}]$, where
$f(\omega)=\{\exp[\omega/k_BT]+1\}^{-1}$ denotes the Fermi distribution.

For the case of symmetric couplings to external electrodes  ($\gamma_L=\gamma_R \equiv \gamma$) and zero temperature ($T=0$), the thermally averaged quantities are given by the following explicit expressions
\begin{align}
\langle c_{1\uparrow }^{\dagger }c_{1\uparrow }\rangle =&\frac{1}{2}- \frac{1}{2\pi}\left(\frac{\epsilon}{E_d}-\frac{\delta}{ E_{t}}\right) \arctan\left(\frac{E_d-E_{t}}{\gamma }\right)\nonumber\\
& -\frac{1}{2\pi}\left(\frac{\epsilon}{E_d}+\frac{\delta}{ E_{t}}\right) \arctan\left(\frac{E_d+E_{t}}{\gamma }\right) ,\label{eqo1}\\
\langle c_{2\downarrow }^{\dagger }c_{2\downarrow }\rangle =&
\frac{1}{2}- \frac{1}{2\pi}\left(\frac{\epsilon}{E_d}+\frac{\delta}{ E_{t}}\right) \arctan\left(\frac{E_d-E_{t}}{\gamma }\right)\nonumber\\
& -\frac{1}{2\pi}\left(\frac{\epsilon}{E_d}-\frac{\delta}{ E_{t}}\right) \arctan\left(\frac{E_d+E_{t}}{\gamma }\right)
,\\
\langle c_{2\downarrow }^{\dagger }c_{1\uparrow }\rangle =&-\frac{t_{so}}{2\pi E_{t}} \left[\arctan\left(\frac{E_d-E_{t}}{\gamma }\right)
\right.\nonumber\\
&\qquad\qquad
\left.
-\arctan\left(\frac{E_d+E_{t}}{\gamma }\right)\right],\\
\langle c_{2\downarrow}c_{1\uparrow }\rangle = &-\frac{\Delta_{12} }{2\pi E_d} \left[\arctan\left(\frac{E_d-E_{t}}{\gamma }\right)
\right.\nonumber\\
&\qquad\qquad\left. +\arctan\left(\frac{E_d+E_{t}}{\gamma }\right)\right] . \label{eqo4}
\end{align}
We have introduced the abbreviations $E_{d}=\sqrt{\epsilon^2+\Delta_{12}^2}$ and $E_{t}=\sqrt{\delta^2+t_{so}^2}$, where the quantum dot energy levels are factorized through $\epsilon _{1\uparrow }=\epsilon+\delta$, and $\epsilon _{2\downarrow }=\epsilon-\delta$. Thus $\epsilon$ is the average energy level and  $2\delta$ denotes their difference.

\begin{figure}
\includegraphics[width=1.0\linewidth,clip]{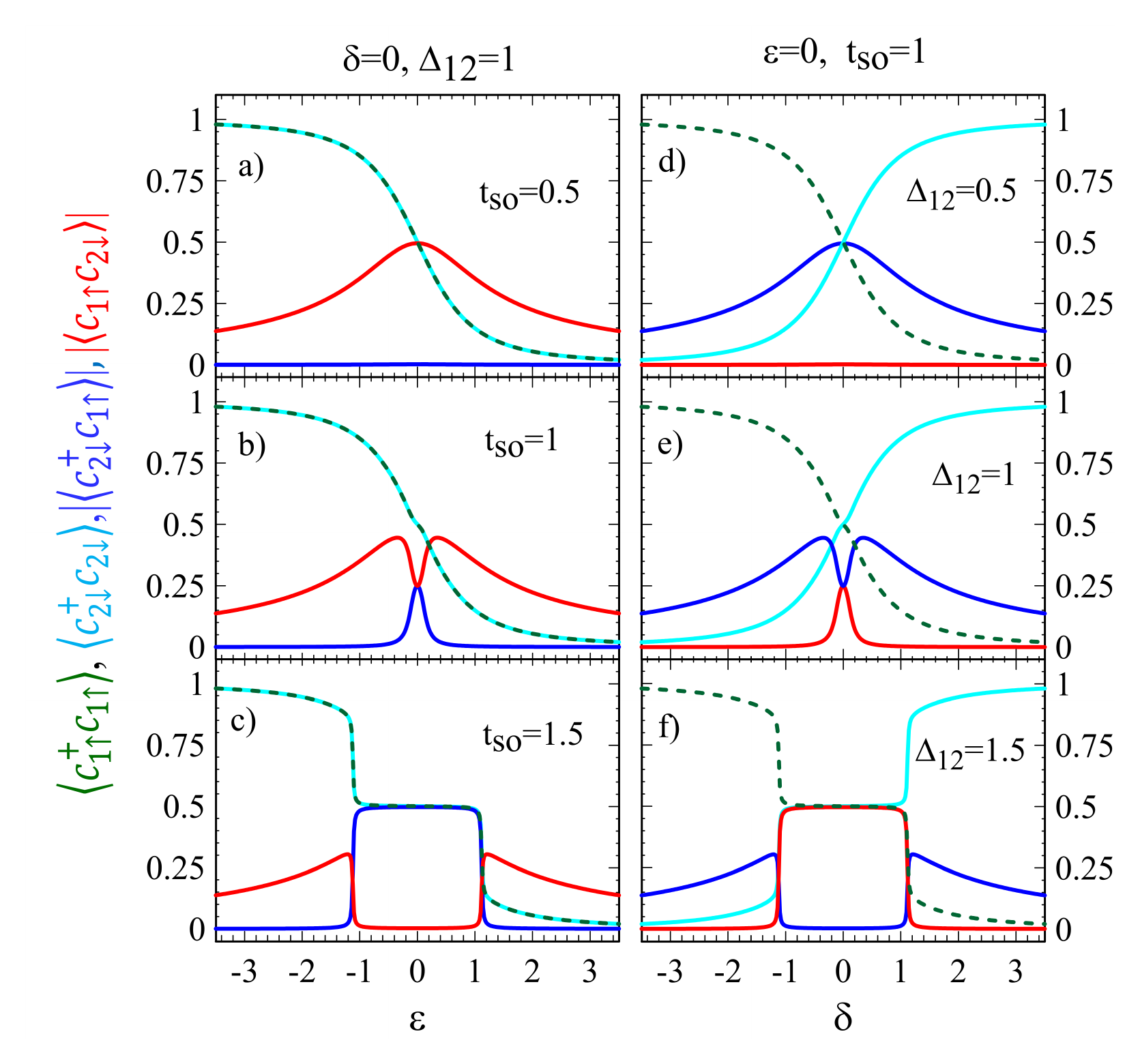}
\caption{Plots of the thermal averages $\langle c^{\dag}_{1\uparrow} c_{1\uparrow}\rangle$ - green dashed, $\langle c^{\dag}_{2\downarrow} c_{2\downarrow}\rangle$ - cyan, $|\langle c^{\dag}_{2\downarrow} c_{1\uparrow}\rangle|$ - blue and $|\langle c_{1\uparrow} c_{2\downarrow}\rangle|$ - red as a function of $\epsilon$  for $\delta=0$, $\Delta_{12}=1$ and  $t_{so}=0.5$, 1, 1.5 (the left column); and as a function of $\delta$ for $\epsilon=0$, $t_{so}=1$ and  $\Delta_{12}=0.5$, 1, 1.5 (the right column). Results are obtained for a small coupling  $\gamma=0.01$ at zero temperature $T=0$.
}\label{forder}
\end{figure}

Figure \ref{forder} presents the thermal averages calculated from Eqs.~(\ref{eqo1}-\ref{eqo4}) for representative sets of the model parameters. In the left column we show the results obtained for the fixed interdot coupling $\Delta_{12}=1$ and three different values of $t_{so}$. Specifically, we plot variation of the expectation values (\ref{eqo1}-\ref{eqo4}) with respect to the energy $\epsilon$, assuming $\delta=0$. In the right column we present the same quantities plotted versus $\delta$, assuming the spin-orbit coupling $t_{so}=1$, $\varepsilon=0$ for three different values of $\Delta_{12}$.

Fig.~\ref{forder}a  shows the results for $t_{so}=0.5<\Delta_{12}$, when the superconducting pairing dominates over the spin-flip processes. In this case, the order parameter $|\langle c_{1\uparrow}c_{2\downarrow} \rangle|$ reaches its optimal value $1/2$ at $\epsilon=0$, whereas the {SFH} order parameter practically vanishes $|\langle c^{\dag}_{2\downarrow} c_{1\uparrow}\rangle|\approx 0$. Fig.~\ref{forder}b shows the results obtained for $t_{so}=\Delta_{12}=1$. We notice the appearance of the SFH order parameter in the small energy region around $\varepsilon=0$, at expense of reducing the superconducting order parameter. At $\epsilon=0$, they both approach the same value $1/4$.
For $t_{so}=1.5>\Delta_{12}=1$, the {CAR} order parameter is much strongly suppressed in the central energy region, whereas the SFH order parameter has its constant value $1/2$. In this particular energy region the quantum dots are half-filled, $n_{1\uparrow}=0.5=n_{2\downarrow}$.

Analogous tendency is presented in the right column of Fig.~\ref{forder}, where the thermal averages (\ref{eqo1}-\ref{eqo4}) are varied against $\delta$ for fixed $t_{so}=1$, $\epsilon=0$, and three different values of $\Delta_{12}$. Now, the order parameters $|\langle c_{1\uparrow}c_{2\downarrow} \rangle|$ and $|\langle c^{\dag}_{2\downarrow} c_{1\uparrow}\rangle|$ exchanged their roles.
Furthermore, we notice anti-symmetry between the average number of $\uparrow$-spin electrons on QD$_1$ 1 and $\downarrow$-spin electrons on QD$_2$ versus the detuning parameter $\delta$.
Besides such antisymmetric occupancy, we again notice clear signatures of a competition between the superconducting and spin-orbit order parameters.
They eventually coexist in a narrow region of the model parameters for the comparable
magnitudes $\Delta_{12} \approx t_{so}$ (see middle panels of Fig.~\ref{forder}). On the other hand, their coexistence is crucial for emergence of the Majorana quasiparticles. Otherwise, the order parameters tend to exclude each other (both when $\Delta_{12}> t_{so}$ and $\Delta_{12} < t_{so}$).
The boundary between these two different quantum phases
occurs at $E_d=E_{t}$. Our results presented in Fig.~\ref{forder} refer to the limit of infinitesimal $\gamma$,  but at larger couplings to external electrodes, the crossover region might undergo modifications accompanied by suppressing the lifetimes of the Majorana quasiparticles.

\section{Currents and Transmission}
\label{currents}

Our final step is devoted to analysis of the key experimentally measurable device characteristics, such as the charge conductance. To this end, we use the Keldysh approach to determine the charge flowing through the biased junction. The operator of charge current transmitted from the normal $L$-th electrode to the first quantum dot is given by
\begin{align}
\hat{I}_{L} =\imath e \;t_{L}\frac{1}{2}\sum_{k,\sigma}(d^{\dag}_{L k\sigma}\tau_z d_{1\sigma}-d^{\dag}_{1\sigma}\tau_z d_{L k\sigma}),
\end{align}
where the operators $d^{\dag}_{L k\sigma}=[c^{\dag}_{Lk\sigma},c_{Lk\sigma}]$ and $d^{\dag}_{1\sigma}=[c^{\dag}_{1\sigma},c_{1\sigma}]$. We introduced the coefficient 1/2 to avoid a double-counting.  Using the lesser Green functions, we can determine expectation value of the spin-resolved current for electrons
\begin{align}\label{eq:current}
I_{Le\sigma} &=\frac{ e }{2h}t_{L}\int d\omega \left[\ll c_{1\sigma}|c^{\dag}_{L\sigma}\gg^{<}_{\omega}-\ll c_{L\sigma}|c^{\dag}_{1\sigma}\gg^{<}_{\omega}\right] .
\end{align}
Similarly we can determine the current for holes, $I_{Lh\sigma}$, as well as the currents from the right electrode, $I_{Re\sigma}$, $I_{Rh\sigma}$.

We restrict our considerations to the $\Psi$ subspace and assume the symmetric coupling, $t_L=t_R$ (i.e. $\gamma_L=\gamma_R=\gamma$), to avoid charge imbalance on the quantum dots, which would  obscure the description of the transport picture. Specifically, we determine the thermally averaged currents $I_{Le\uparrow}$, $I_{Lh\uparrow}$, $I_{Re\downarrow}$ and $I_{Re\downarrow}$, using Eq.~(\ref{eq:current}).
From the symmetry relations $(I_{Le\uparrow }-I_{Lh\uparrow})+(I_{Re\downarrow}-I_{Rh\downarrow })=0$ and $(I_{Le\uparrow }-I_{Lh\uparrow})-(I_{Re\downarrow}-I_{Rh\downarrow })=0$ (see Appendix~\ref{conservation} for details) we infer the current conservation between the electron and hole channels~\cite{Chamon2010}. We can next determine $I_{S}=(I_{Le\uparrow}+I_{Lh\uparrow})+(I_{Re\downarrow}+I_{Rh\downarrow})$ and $I_{LR}=[(I_{Le\uparrow}+I_{Lh\uparrow})-(I_{Re\downarrow}+I_{Rh\downarrow})]/2$. Since these functions are given in analytical form, we can regroup their integrands with respect to the distribution functions
$F_S=(f_{Le}-f_{Rh})+(f_{Re}-f_{Lh})$, $F_{LR}=(f_{Le}-f_{Re})+(f_{Rh}-f_{Lh})$ and $F_{eh}=(f_{Le}-f_{Re})-(f_{Rh}-f_{Lh})$. Finally, we obtain the compact expressions
\begin{align}
I_S=&
-\frac{e}{h}\int d\omega\; [(f_{Le}-f_{Rh})+(f_{Re}-f_{Lh})] T_S(\omega)\nonumber\\
&-\frac{e}{h}\int d\omega\; [(f_{Le}-f_{Re})-(f_{Rh}-f_{Lh})] T^{eh}_{S}(\omega),\label{IS}\\
I_{LR}=&
-\frac{e}{2h}\int d\omega\;[(f_{Le}-f_{Re})+(f_{Rh}-f_{Lh})] T_{LR}(\omega)\nonumber\\
&-\frac{e}{2h}\int d\omega\; [(f_{Le}-f_{Re})-(f_{Rh}-f_{Lh})] T^{eh}_{LR}(\omega),\label{ILR}
\end{align}
for the current flowing from/to the  superconductor and the other current between the normal electrodes, respectively. The transmission coefficients can be derived for the arbitrary set of model parameters $\epsilon$, $\delta$, $t_{so}$, $\Delta_{12}$ and are given by
\begin{align}\label{ts}
&T_S(\omega)=4\gamma^2 \Delta_{12}^2\{\omega^4+2\omega^2(3E_t^2-E_d^2+\gamma^2)\nonumber\\
&\qquad  + [(E_t-E_d)^2+\gamma ^2] [(E_t-E_d)^2+\gamma^2]\}/m,\\
\label{tseh}
&T^{eh}_S(\omega)= 16\delta \gamma^2 \Delta_{12}^2  \omega(\omega^2+E_t^2-E_d^2+\gamma^2)/m,\\
&T_{LR}(\omega)= 4\gamma^2 t_{so}^2 \{\omega^4+2\omega^2(3E_d^2-E_t^2+\gamma^2)\nonumber\\
&\qquad  +
[(E_t-E_d)^2+\gamma ^2] [(E_t-E_d)^2+\gamma^2]
\}/m,\label{tlr}\\
&T^{eh}_{LR}(\omega)=16 \epsilon \gamma^2 t_{so}^2 \omega(\omega^2-E_t^2+E_d^2+\gamma^2)/m,
\label{tlreh}
\end{align}
where the denominator $m=[(\omega+E_t-E_d)^2+\gamma^2]
[(\omega-E_t+E_d)^2+\gamma^2] [(\omega-E_t-E_d)^2+\gamma^2] [(\omega+E_t+E_d)^2+\gamma^2]$. The coefficient $T_S$ describes transport of the Cooper pairs from superconductor, whereas the term $T^{eh}_S$ appears only when particle-hole symmetry is absent. Similarly, $T_{LR}$ and $T^{eh}_{LR}$ describe the symmetric and antisymmetric (in $\omega$) contributions of electrons and holes to the charge current between the normal electrodes.
Notice that $T_{S}$ and $T_{LR}$  are dependent only on $E_d$ and $E_t$ and they neither depend on $\epsilon$ nor on $\delta$. Furthermore, the transmission coefficients of the S and  LR sectors, Eqs.~(\ref{ts}-\ref{tlreh}), are dual upon exchanging the appropriate model parameters.

There are various ways to apply the bias voltage to our system. One can apply a symmetric bias $\mu_L=eV/2$, $\mu_R=-eV/2$, for which the differential conductances $\mathcal{G}_{LR}\equiv d I_{LR}/d V$ and $\mathcal{G}_S\equiv d I_{S}/d V$ at zero temperature are given by
\begin{align}\label{glr}
\mathcal{G}_{LR}&
 =\frac{e^2}{2h} [T_{LR}(eV/2)+T_{LR}(-eV/2)],\\
 \label{gs}
 \mathcal{ G}_S&=0.
\end{align}
For the Cooper pair splitter biasing, $\mu_L=eV=\mu_R$, one gets
\begin{align}
\label{glrsp}
\mathcal{ G}_{LR}&=0,\\
\label{gssp}
\mathcal{G}_S
&=\frac{e^2}{h}[2T_S(eV)+2T_S(-eV)].
\end{align}
For asymmetric bias, $\mu_L=eV$, $\mu_R=0$, the situation is more complex because the differential conductances
\begin{align}
\mathcal{ G}_{LR}&=\frac{e^2}{2h}[T_{LR}(eV)+T_{LR}(-eV)\nonumber\\
&\qquad\qquad\qquad+ T^{eh}_{LR}(eV) - T^{eh}_{LR}(-eV)] ,\\
\mathcal{G}_S&=\frac{e^2}{h}[T_S(eV)+T_S(-eV)+T^{eh}_S(eV)-T^{eh}_S(-eV)]
\end{align}
contain then the terms with $T^{eh}_{LR}$ and $T^{eh}_{S}$, describing asymmetric contributions of electrons and holes to the currents.
To calculate the net current one has to collect the contributions from  the  $\Psi$ and $\Phi$ subspaces.

\begin{figure}
\includegraphics[width=1\linewidth,clip]{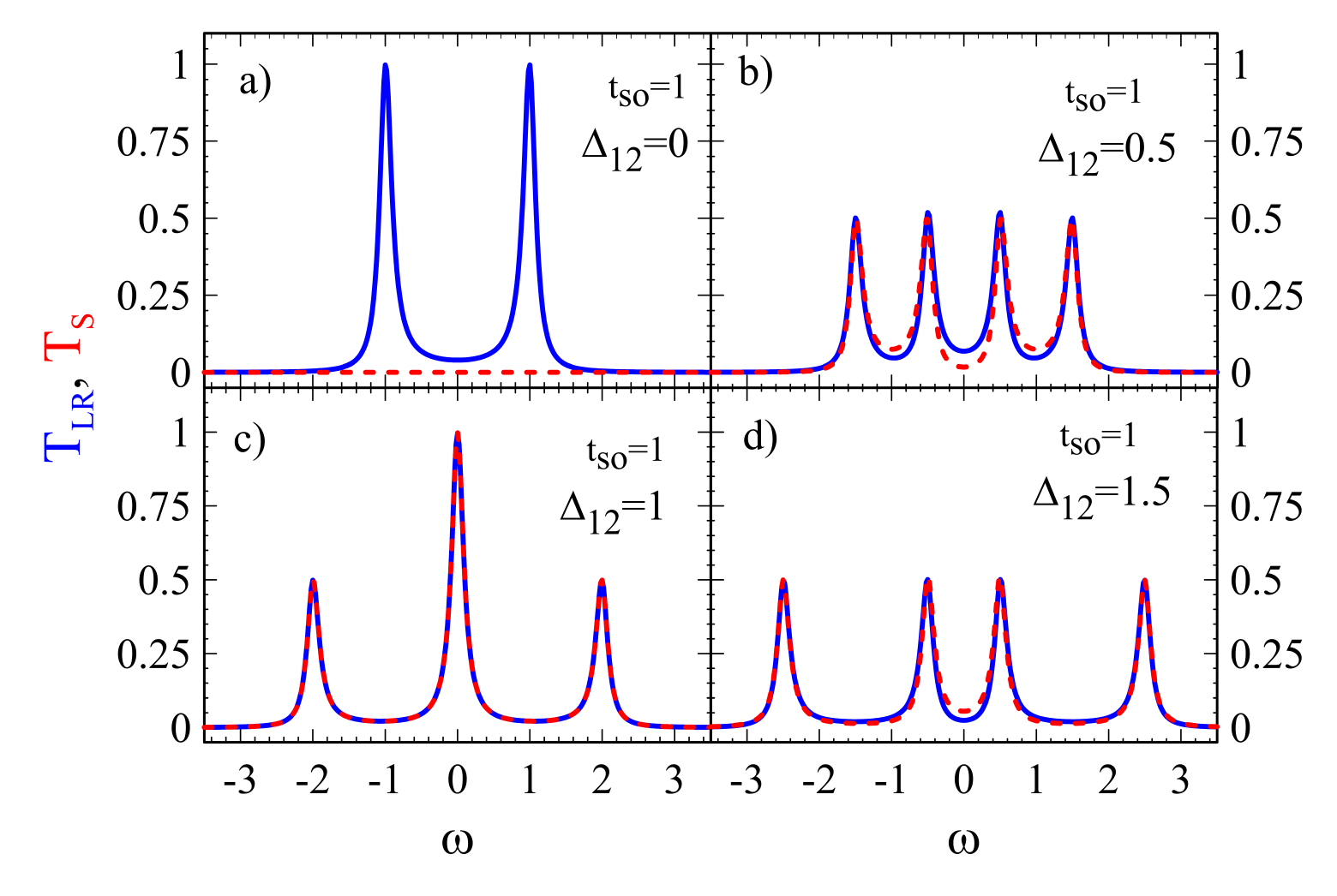}

\caption{Evolution of the transmission $T_{LR}(\omega)$ (blue) and $T_{S}(\omega)$ (dashed red) for $t_{so}=1$  and several $\Delta_{12}=0$, 0.5, 1 and 1.5 at $\epsilon_1=0$, $\epsilon_2=0$ and a small symmetric coupling $\gamma=0.1$ to the reservoirs. Compare with the spectral density in Fig.\ref{Fspectral}.}\label{transm1}
\end{figure}

\begin{figure}
\includegraphics[width=1\linewidth,clip]{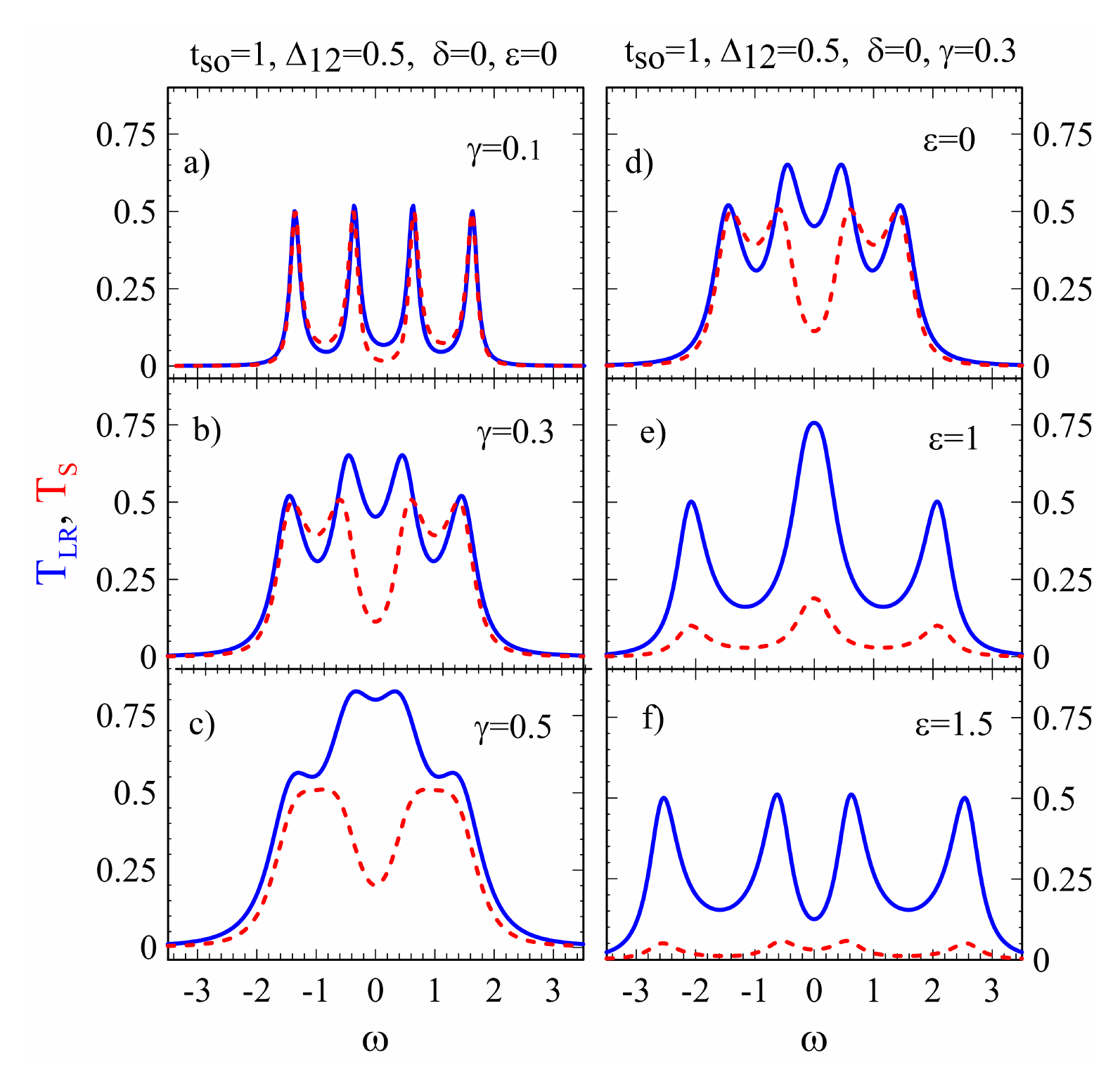}
\caption{Left-column (figure a,b and c) presents transmission $T_{LR}(\omega)$ (blue) and $T_{S}(\omega)$ (dashed-red) for various couplings $\gamma=0.1$, $\gamma=0.3$ and $\gamma=0.5$ at $\epsilon=0$ and  $\delta=0$. Right column (figure d,e,and f) presents how $T_{LR}(\omega)$ and $T_S(\omega)$ change for various $\epsilon=0$, 1 and 1.5 at $\gamma=0.3$ and $\delta=0$. The other parameters are $t_{so}=1$ and $\Delta_{12}=0.5$}\label{transm2}
\end{figure}

Let us analyze the transmission $T_{LR}$ and $T_{S}$, Eqs.(\ref{tlr}) and (\ref{ts}). These quantities can be determined in measurements of a differential conductance applying the symmetric bias voltage and in the CPS bias configuration, respectively.
Fig.~\ref{transm1} presents evolution of the transmission $T_{LR}$ and $T_{S}$  for $t_{so}=1$  and several values of $\Delta_{12}$ at small $\gamma=0.1$. The plots resemble the spectral density shown in Fig.~\ref{Fspectral}. For $t_{so}=1$ and $\Delta_{12}=0$ there are two quasiparticle states, therefore $T_{LR}$  reveals two peaks, while $T_{S}=0$. Switching on $\Delta_{12}$ leads to a splitting of the transmission into four-peak structure, each of them containing half of the initial spectral weight. At $\Delta_{12}=t_{so}=1$  two internal peaks merge, forming the central peak of height with the doubled intensity. For $\Delta_{12}>1$ the transmittance is again characterized by the four peak-structure.

Let us focus on the zero-energy state, appearing in the spectral density (in Fig.~\ref{Fspectral}) which corresponds to the Majorana quasiparticles spatially separated on different quantum dots [see Eq.~(\ref{GFAM})].  Fig.~\ref{transm1}c shows the transmission coefficient at such {\em sweet spot}, revealing the zero-energy enhancement of $T_{LR}(\omega=0)\approx1$.  Moreover, we observe clear evidence for the duality
\begin{align}
T_{LR}(\omega)&=T_S(\omega)=
\frac{\gamma ^2 q}{2(q^2+\gamma^2)}
\left[\frac{2 q}{\omega^2+\gamma ^2} \right.
\nonumber\\ &\qquad \left.
-\frac{\omega-3 q}{(\omega-2 q)^2+\gamma ^2}+\frac{\omega +3 q}{(\omega+2 q)^2+\gamma ^2}\right]
,
\end{align}
where $t_{so}=\Delta_{12}\equiv q$ and $\epsilon_1=\epsilon_2=0$. Here, we have performed the spectral decomposition in order to show the contribution of each pole. This result means that the indirect transport between the normal L and R-electrodes for the symmetric bias is fully equivalent to the transport of entangled electrons in the CPS configuration. This case shows how prefect interplay {SFH} and CAR
enhances the transmission at zero voltage $V\rightarrow 0$ \cite{Prada2020, Danon2020,Pan2021}. In what follows, we show in the strong coupling limit $\gamma$ that these features are destroyed by interference and dissipation processes.

In contrast to the spectral density, the transmission peaks are not Lorentzian, and there is no simple line shape of the resonances. In general,  $T_{LR}$ is distinct from $T_S$ because they are affected by different quantum interference processes. This is particularly evident for the stronger coupling $\gamma$ presented in Fig.~\ref{transm2}a-c, where {SFH} dominates over CAR $t_{so}=2\Delta_{12}$ (for comparison see Fig.\ref{forder} and the discussion in Sec. \ref{equilibrium}). In this strong coupling case, the transmittance $T_{LR}$ is strongly enhanced in a central part of the plot, whereas $T_{S}$ is far less sensitive to $\gamma$.

Figures \ref{transm2}d-f present the transmissions for various values of $\epsilon$. This parameter shifts the energy levels of quantum dots, where the CAR order parameter weakens (see Fig.\ref{forder}). Such influence is well noticeable in the plot of $T_{S}$, while $T_{LR}$ is much less affected. In particular, at $\epsilon=1$ there appears the quasiparticle state, responsible for the crossover behavior presented in Fig.~\ref{forder}. In this case one can notice the zero-energy state, however, it does not refer to the true Majorana state because particle-hole symmetry is missing (this important criterion has been also stressed in Ref.~\cite{Luethi2024}). Moreover, the transmission coefficients $T_{LR}$ and $T_S$ are quite different, indicating the low electron entanglement.

\section{Conclusions}
\label{concl}

We have analyzed the spectral and transport properties of the sandwich-like setup, comprising two quantum dots placed between s-wave superconductor and a semiconductor with the strong SOI. We have focused our investigations on the influence of the spin-flip hopping between the quantum dots on the molecular Andreev states, originating from the superconducting proximity effect. We have analyzed systematic  evolution of these molecular bound states with respect to the varying model parameters, and we have explicitly shown under what conditions (at the {\em sweet spot}) the Majorana-like states could be realized. In the present setup, they represent the zero-energy quasiparticles  which are fully spin-polarized and localized on different quantum dots.

Sandwich-type architecture proposed in this work offers
relatively easy control of the Andreev and spin-orbit processes, which give rise to formation of the zero-energy states at two QDs. Furthermore, we have proved that the Andreev reflection and the spin-exchange processes are dual. This duality is well manifested in the charge currents: for the symmetric bias voltage and in the CPS bias voltage configuration, Eq.(\ref{glr})-(\ref{gs}) and (\ref{glrsp})-(\ref{gssp}), respectively.
We stress that the superconducting electrode plays an active role in transport for both bias configurations. These transport processes are characterized by the transmission coefficients $T_{LR}$ and $T_{S}$, Eq.(\ref{tlr})-(\ref{tseh}), which are sensitive to interplay of the interdot electron pairing with the spin-orbit order.

In particular, we have shown how the Andreev bound states are hybridized with the spin-orbit bonding states for particle-hole symmetry, inducing four Dirac states. Specifically, at the sweet spot, $\Delta_{12}=t_{so}=g$, we predict emergence of the Majorana quasiparticle states. They are degenerate zero-energy modes which in our setup are characterized by opposite spin sectors and exist on different quantum dots, being coupled only to their own reservoirs Eq.~(\ref{GFAM}). The zero-energy quasiparticles are manifested by an (almost) perfect transmission, with $T_{LR}(0)=T_S(0)\approx 1$. We predict the duality to be observable by the transmission coefficients $T_{LR}(\omega)=T_S(\omega)$ for arbitrary $\omega$, originating from perfect entanglement of the transferred electrons. Dissipation to the external normal electrodes (large $\gamma$), however, can destroy these features. To verify our predictions, we suggest performing conductance measurements for both configurations of the applied bias voltage.

The key issue which remains to be addressed, is some experimental realization of this hybrid sandwich system. It would require two QDs to be strongly proximized by superconductor on one face and in parallel situated to a semiconductor on the opposite face.  We hope that our theoretical predictions might encourage experimentalists to undertake this task.

\acknowledgments
T.D. and K.I.W. acknowledge support by the National Science Centre, Poland within the Weave-Unisono programme through grant no. 2022/04/Y/ST3/00061.

\appendix
\begin{widetext}
\section{Current conservation and their symmetry}
\label{conservation}

Let us consider the currents
$(I_{Le\uparrow }-I_{Lh\uparrow}) +(I_{Re\downarrow}-I_{Rh\downarrow })$. For the symmetric coupling, $t_L=t_R$, we  calculated all lesser Green functions and can prove that the corresponding integrand is
\begin{align}
&\left[\left(\ll c_{1\uparrow}|c^{\dag}_{L\uparrow}\gg^{<}_{\omega}-\ll c_{L\uparrow}|c^{\dag}_{1\uparrow}\gg^{<}_{\omega}\right)-\left(\ll c^{\dag}_{1\uparrow}|c_{L\uparrow}\gg^{<}_{\omega}-\ll c_{L\uparrow}|c^{\dag}_{1\uparrow}\gg^{<}_{\omega}\right)\right]\nonumber\\
&+\left[\left(\ll
c_{2\downarrow}|c^{\dag}_{R\downarrow}\gg^{<}_{\omega}-\ll c_{R\downarrow}|c^{\dag}_{2\downarrow}\gg^{<}_{\omega}\right)-\left(\ll c^{\dag}_{2\downarrow}|c_{R\downarrow}\gg^{<}_{\omega}-\ll c_{R\downarrow}|c^{\dag}_{2\downarrow}\gg^{<}_{\omega}\right)\right]=0,
\end{align}
for any model parameters. This is equivalent to the current conservation rule. Thus, the current averages
$(I_{Le\uparrow }-I_{Lh\uparrow}) +(I_{Re\downarrow}-I_{Rh\downarrow })=0$.

For the other current averages we group their integrands with respect to the distribution functions
$F_{LR}=(f_{Le}-f_{Re})+(f_{Rh}-f_{Lh})$, $F_S=(f_{Le}-f_{Rh})+(f_{Re}-f_{Lh})$ and $F_{eh}=(f_{Le}-f_{Re})-(f_{Rh}-f_{Lh})$. Next, using the symmetry of the integrands, one can show
\begin{align}\label{flr}
\int_{-\infty}^{\infty} d\omega\; F_{LR}\;g_{LR}(\omega)&=
\int_{-\infty}^{\infty} d\omega\; (f_{Le}-f_{Re})[g_{LR}(\omega)+g_{LR}(-\omega)],\\
\label{fs}
\int_{-\infty}^{\infty} d\omega\; F_S\; g_S(\omega)&=
\int_{-\infty}^{\infty} d\omega\; (f_{Le}-f_{Rh})[g_S(\omega)+g_S(-\omega)],\\
\label{feh}
\int_{-\infty}^{\infty} d\omega F_{eh}\; g_{eh}(\omega)&=
\int_{-\infty}^{\infty} d\omega\; (f_{Le}-f_{Re})[g_{eh}(\omega)-g_{eh}(-\omega)],
\end{align}
where $g_{LR}(\omega)$, $g_S(\omega)$ and $g_{eh}(\omega)$ are some frequency-dependent functions obtained from the lesser Green functions. Let us first consider
\begin{align}
\left(I_{Le\uparrow }-I_{Lh\uparrow }\right) -&\left(I_{Re\downarrow}-I_{Rh\downarrow}\right)
=\nonumber\\
& \frac{e}{h}\int_{-\infty}^{\infty} d\omega\;
8 \gamma ^2 \left\{4F_{LR} \omega \epsilon t_{so}^2 \left[\omega^2-E_t^2+E_d^2+\gamma ^2\right]+4F_S \omega \delta \Delta_{12}\left[\omega^2+E_t^2-E_d^2+\gamma ^2\right]\right.\nonumber\\
&+F_{eh}\left[\omega^4(\Delta_{12}^2+t_{so}^2)+2 \omega^2(-4\Delta_{12}^2 t_{so}^2-E_d^2(\Delta_{12}^2-3 t_{so}^2)+E_t^2 (3\Delta_{12}^2-t_{so}^2) +(\Delta_{12}^2+t_{so}^2)\gamma^2)\right. \nonumber\\
&+\left.\left.(\Delta_{12}^2+t_{so}^2)((E_t^2-E_d^2)^2+2(E_t^2+E_d^2)\gamma^2+\gamma^4) -8\Delta_{12}^2 t_{so}^2\gamma^2 \right]\right\}/m.
\end{align}
In this case the corresponding functions $g_{LR}(\omega)$ and $g_S(\omega)$ are odd, while $g_{eh}(\omega)$ is even with respect to $\omega$. Thus, using the relations (\ref{flr})-(\ref{feh}) one gets
 $\left(I_{Le\uparrow }-I_{Lh\uparrow }\right) -\left(I_{Re\downarrow}-I_{Rh\downarrow}\right)
= 0$.  Similarly, we determine the average currents $I_{S}$ and $I_{LR}$, Eqs.(\ref{IS})-(\ref{ILR}), with the corresponding transmission coefficients (\ref{ts})-(\ref{tlreh}).
\end{widetext}

\bibliography{splitterSO}
\end{document}